\documentclass[fleqn,11pt]{article}
\usepackage{amsmath,amssymb,amsfonts,graphicx}

\usepackage{hyperref}

\usepackage{amsfonts,amssymb,cite}

\def\noi{\noindent}

\newcommand{\Title}[1]{\noi {{\Large\bf #1}}\\[1ex]}

\def\Aunames#1{\noi{\bf #1}}
\def\au#1{${}^{#1}$}
\def\Addresses#1{\medskip\noi \protect
	\begin{description}\itemsep -3pt {\it #1} \end{description}}
\def\adr#1#2{\item[${}^{#1}$]{\it #2}}

\newcommand{\Abstract}[1]{\vskip 2mm \begin{center}
        \parbox{16.4cm}{\small\noi #1} \end{center}\medskip}

\def\email#1#2{\footnotetext[#1]{e-mail: #2}\addtocounter{footnote}{1}}

\def\nq{\hspace*{-1em}}
\def\nqq{\hspace*{-2em}}
\def\nhq{\hspace*{-0.5em}}

\def\cm{\hspace*{1cm}}
\def\inch{\hspace*{1in}}

\def\Jl#1#2{#1 {\bf #2},\ }

\def\ApJ#1 {\Jl{Astroph. J.}{#1}}
\def\CQG#1 {\Jl{Class. Quantum Grav.}{#1}}
\def\DAN#1 {\Jl{Dokl. AN SSSR}{#1}}
\def\GC#1 {\Jl{Grav. Cosmol.}{#1}}
\def\GRG#1 {\Jl{Gen. Rel. Grav.}{#1}}
\def\JETF#1 {\Jl{Zh. Eksp. Teor. Fiz.}{#1}}
\def\JETP#1 {\Jl{Sov. Phys. JETP}{#1}}
\def\JHEP#1 {\Jl{JHEP}{#1}}
\def\JMP#1 {\Jl{J. Math. Phys.}{#1}}
\def\NPB#1 {\Jl{Nucl. Phys. B}{#1}}
\def\NP#1 {\Jl{Nucl. Phys.}{#1}}
\def\PLA#1 {\Jl{Phys. Lett. A}{#1}}
\def\PLB#1 {\Jl{Phys. Lett. B}{#1}}
\def\PRD#1 {\Jl{Phys. Rev. D}{#1}}
\def\PRL#1 {\Jl{Phys. Rev. Lett.}{#1}}

\def\lal{&&\nqq {}}
\def\eq{Eq.\,}
\def\eqs{Eqs.\,}
\def\beq{\begin{equation}}
\def\eeq{\end{equation}}
\def\bear{\begin{eqnarray}}
\def\bearr{\begin{eqnarray} \lal}
\def\ear{\end{eqnarray}}
\def\earn{\nonumber \end{eqnarray}}

\def\nnn{\nonumber\\ \lal }

\def\yy{\\[5pt] {}}
\def\yyy{\\[5pt] \lal }

\def\tst{\textstyle}

\def\fract#1#2{{\tst\frac{#1}{#2}}}

\def\half{{\fract{1}{2}}}

\def\e{{\,\rm e}}
\def\d{\partial}

\def\sign{\mathop{\rm sign}\nolimits}
\def\diag{\mathop{\rm diag}\nolimits}

\def\const{{\rm const}}
\def\eps{\varepsilon}

\def\then{\ \Rightarrow\ }
\newcommand{\toas}{\mathop {\ \longrightarrow\ }\limits }

\newcommand{\vars}[1]{\left\{\begin{array}{ll}#1\end{array}\right.}

\def\eqn#1{\eq\eqref{#1}}
\def\rf{\eqref}
\def\arccot{\mathop{\rm arccot}\nolimits}

\def\mn{_{\mu\nu}}
\def\MN{^{\mu\nu}}
\def\mN{_\mu^\nu}

\def\kappa{\varkappa}

\def\hG{{\hat\Gamma}}

\def\R{{\mathbb R}}

\def\og{{\bar g}}
\def\ophi{{\bar\phi}}
\def\oG{{\bar G}}
\def\oR{{\bar R}}
\def\oT{{\bar T}}

\def\ME{\mathbb{M}_{\rm E}}
\def\MJ{\mathbb{M}_{\rm J}}

\def\ssph{static, spherically symmetric}

\def\bh{black hole}
\def\bhs{black holes}
\def\wh{wormhole}
\def\whs{wormholes}
\def\asflat{asymptotically flat} 
\def\Scw{Schwarz\-schild}
\def\RN{Reiss\-ner-Nord\-str\"om}

\usepackage{color}

\begin{document}
\thispagestyle{empty}
\twocolumn[

\bigskip

\Title {Spherically symmetric space-times in generalized hybrid\yy
		 metric-Palatini gravity}
	
\Aunames{K. A. Bronnikov,\au{a,b,c1} S. V. Bolokhov,\au{b;2} and M. V. Skvortsova\au{b;3}}

\Addresses{\small
\adr a {Center for Gravitation and Fundamental Metrology, VNIIMS, 
		Ozyornaya ul. 46, Moscow 119361, Russia}
\adr b {Institute of Gravitation and Cosmology, 
		Peoples' Friendship University of Russia (RUDN University),\\ 
		ul. Miklukho-Maklaya 6, Moscow 117198, Russia}
\adr c {National Research Nuclear University ``MEPhI'', 
		Kashirskoe sh. 31, Moscow 115409, Russia}
        }
                
\Abstract
   {We discuss vacuum static, spherically symmetric asymptotically flat solutions of the generalized
   hybrid metric-Palatini theory of gravity (generalized HMPG) suggested by B\"ohmer and Tamanini,  
   involving both a metric $g\mn$ and an independent connection $\hG\mn^\alpha$; the gravitational 
   field Lagrangian is an arbitrary function $f(R,P)$ of two Ricci scalars, $R$ obtained from $g\mn$ 
   and $P$ obtained from $\hG\mn^\alpha$. The theory admits a scalar-tensor representation with 
   two scalars $\phi$ and $\xi$ and a potential $V(\phi,\xi)$ whose form depends on $f(R,P)$. 
   Solutions are obtained in the Einstein frame and transferred back to the original Jordan frame 
   for a proper interpretation. In the completely studied case $V \equiv 0$, generic solutions 
   contain naked singularities or describe traversable wormholes, and only some special cases 
   represent black holes with extremal horizons. For $V(\phi,\xi) \ne 0$, some examples of 
   analytical solutions are obtained and shown to possess naked singularities. Even in the cases 
   where the Einstein-frame metric $g^E\mn$ is found analytically, the scalar field equations need a 
   numerical study, and if $g^E\mn$ contains a horizon, in the Jordan frame it turns to a singularity
   due to the corresponding conformal factor. }
\bigskip

 ] 
\email 1 {kb20@yandex.ru} 
\email 2 {boloh@rambler.ru}
\email 3 {milenas577@mail.ru}

\section{Introduction}

   The century-old general relativity (GR) still amazingly well passes all local gravitational tests. 
   Nevertheless, hundreds of alternative theories of gravity are under consideration. Motivations for these 
   studies are both theoretical and empirical \cite{DE,ex-GR1, ex-GR2}. Theoretical difficulties of GR
   include the problems with its quantization and the existence of space-time singularities in the most
   relevant solutions of the theory. The main empirical difficulty of GR is its inability to account for 
   extra gravitating matter in galaxies and the accelerated expansion of the Universe (the so-called 
   Dark Matter and Dark Energy problems).
   
   The theory called Hybrid metric-Palatini gravity (HMPG), put forward in \cite{har12}, is one of 
   such alternatives. It assumes the Riemannian nature of physical space-time but, along with the
   metric $g\mn$, it postulates the existence of an independent connection $\hG\mn^\alpha$.
   The total action of HMPG reads \cite{har12}
\beq  				\label{S}
	S = \frac{1}{2\kappa^2}\int d^4 x\sqrt{-g} [R + F(P)] + S_m,
\eeq    
   where $R = R[g]$ is the scalar curvature obtained in the usual way from $g\mn$, while $F(P)$ 
   is a function of another Ricci scalar $P = g\MN P\mn$ corresponding to the Ricci tensor $P\mn$ 
   obtained in the standard way from the independent connection $\hG\mn^\alpha$; furthermore, 
   $g = \det(g\mn)$, $ \kappa^2$ is the gravitational constant, and $S_m$ is the action of 
   nongravitational matter.
  
   HMPG has been shown to be in agreement with the classical tests of gravity in the Solar 
   system \cite{cap13b}, and it also fairly well describes the observed dynamics in galaxies 
   and galaxy clusters, thus successfully trying to explain the Dark Matter problem \cite{cap13a}.
   At the cosmological level, it has been shown to create models of accelerated expansion without 
   invoking a cosmological constant \cite{cap13}; for more detailed descriptions see the reviews 
   \cite{cap15,har18} and also \cite{bor15} for a study of Noether symmetries in HMPG, and 
   \cite{edery} for a discussion of a relationship between HMPG and $R^2$ gravity. Spherically 
   symmetric static solutions of HMPG, describing, in particular, \bhs\ and \whs, were studied in
   \cite{dan19,kb19,hyb-sph}, and static cylindrical stringlike objects in \cite{har20, hyb-cy}.
   
   A natural extension of HMPG, treating the curvature scalars $R$ and $P$ on equal grounds and 
   thus introducing an arbitrary function of both $R$ and $P$, has been suggested in \cite{boh13}.
   We will call it, for short, the Generalized Hybrid Theory (GHT).  
   Many results of interest have been obtained in this theory. Thus, cosmological solutions have been 
   obtained and studied in \cite{hyb-cos1, hyb-cos2, sa20}, in particular, it has been shown that 
   this theory makes possible a unified description of Dark Energy and Dark Matter \cite{sa20}.
   The weak-field phenomenology in GHT was studied in \cite{mont19w,rosa21w}. Apart from 
   analyzing the constraints following from the solar-system tests of gravity, gravitational waves 
   in GHT were discussed, and it was concluded \cite{lem20k} that, unlike other theories with 
   scalar modes, in this model the two effective scalar degrees of freedom can interact and 
   produce beatings. J. Rosa et al. \cite{rosa18} found a family of \ssph\ wormhole solutions of
   GHT, in which the matter content respects the Null Energy Condition (NEC), which is well known to 
   be impossible in GR. The same authors  \cite{lem20k} found the conditions under which vacuum 
   solutions of GR with $R\mn =0$, such as the Schwarzschild and Kerr solutions, are also solutions 
   of GHT, and investigated the stability conditions for the Kerr solution in this theory. A review 
   encompassing both the original HMPG and its generalized version can be found in \cite{lobo20-rev}.
   
   The present paper extends our previous study of \ssph\ solutions of HMPG \cite{kb19,hyb-sph}
   to GHT with a Lagrangian depending on both $R$ and $P$. According to 
   \cite{boh13}, this theory, like HMPG as such, has a scalar-tensor representation, but now with 
   two scalar fields, one of which is canonical while the other may be canonical or phantom,
   with a self-interaction potential $V$, whose form depends on the form of the initial function $f(R,P)$
   specifying the particular theory, see \eqs \rf{GS}--\rf{VRP}.
   The interplay of these scalars leads to a wide variety of solutions even in the case of a zero 
   potential $V$ of the two scalars, corresponding to a particular set of the functions  $f(R,P)$,
   depending on a function of a single variable.
   As in HMPG, the case $V=0$ is closely related to solutions with a conformal scalar field in GR. 
   We also present some simple examples of solutions with a nonzero potential, where, even 
   though the metric is found analytically, the scalar field equations need a numerical study.
       
   The paper is organized as follows. In the next section we discuss the main features of the
   STT representation of GHT \cite{boh13}, in particular, a transition to the 
   Einstein conformal frame. In Section 3 we derive and analyze \ssph\ solutions in the simplest 
   case ($V =0$). In Sec. 4 we discuss some examples of solutions with $V \ne 0$, where the 
   Einstein-frame metric is found analytically by analogy with similar solutions of GR, but the 
   scalar field that determines a transition to the Jordan frame (in which the theory is initially
   formulated) is found only numerically. Section 5 is a conclusion.
  
\section {GHT and its scalar-tensor representation}  

   The extended (or generalized) hybrid metric-Palatini gravity theory (GHT) supposes
   that the physical 4D space-time contains a Riemannian metric $g\mn$ and an independent
   connection $\hG\mn^\alpha$. The total action reads \cite{boh13}
\beq  				\label{GS}
			S = \frac{1}{2\kappa^2}\int d^4 x\sqrt{-g} f (R,P) + S_m,
\eeq    
   where $R = R[g]$ is the scalar curvature derived from $g\mn$, $P$ is the scalar 
   $P = g\MN P\mn$ obtained with the Ricci tensor $P\mn$ built in the standard way from 
   the connection $\hG\mn^\alpha$, $g = \det(g\mn)$, $ \kappa^2$ is the gravitational constant, 
   $S_m$ is the action of nongravitational matter, and $f(R,P)$ is an arbitrary smooth function 
   of two variables, subject to certain physical requirements.
  
   Variation of \rf{GS} in the independent connection $\hG\mn^\alpha$ leads to the conclusion 
   \cite{boh13} that $\hG\mn^\alpha$ is the Riemannian (Levi-Civita) connection
   corresponding to a metric conformal to $g\mn$, namely, $h\mn = f_P g\mn$, with the 
   conformal factor $f_P \equiv \d f/\d P$. Furthermore, as shown in \cite{boh13}, under the 
   condition that the Hessian of $f(R,P)$ is nonzero, that is,
\beq 					\label{Hess}
                    f_{RP}^2 \ne f_{RR} f_{PP}
\eeq   
  (the indices $R$ and $P$ denote partial derivatives with respect to $R$ and $P$), 
   the whole theory admits a reformulation as a scalar-tensor theory with two scalar fields
   where the gravitational part of the action is
\beq                           \label{S1} 
	S_g = \! \int \! d^4x \sqrt{-g}\bigg[\chi R - \xi P - 2V(\chi, \xi)\bigg], 
\eeq          
   where\footnote
  		{We safely omit the factor $1/(2\kappa^2)$ at the gravitational part of the action since 
	  	only vacuum configurations, where $S_m =0$, will be considered.} 
\beq                        \label{scalars}
			\chi = f_R, \qquad \xi = - f_P, 
\eeq  	
   and the potential $V(\chi, \xi) $ is related to $f(R, P)$ by
\beq                             			\label{VRP}
		2V(\chi, \xi) = - f + \chi R - \xi P.
\eeq    
    Using the expression of $P$ in terms of $\xi$ and $g\mn$, the action \rf{S1} can be rewritten 
    as (up to boundary terms) \cite{boh13}\footnote
    		{Unlike \cite{har12, cap15, boh13} etc., we are using the metric signature $(+  -   -\, -)$, 
	  	hence the plus sign before $(\d\phi)^2 = g\MN \phi_{,\mu}\phi_{,\nu}$ corresponds to a 
  		canonical field and a minus to a phantom field. The Ricci tensor is defined as 
	  	$R\mn =\d_\nu  \Gamma^\alpha_{\mu\alpha} - \ldots$, so that, for example, the scalar 
	  	curvature is positive in de Sitter space-time. We also use the units in which $c = G = 1$
  		($c$ being the speed of light and $G$ the Newtonian gravitational constant.} 	 
\beq               \label{S2}  		
	S_g = \! \int \! d^4x \sqrt{-g}
				\bigg[\phi R + \frac{3}{2\xi} (\d\xi)^2 - 2W(\phi, \xi)\bigg], 
\eeq
   where we have introduced the new scalar field $\phi \equiv \chi - \xi$, and 
   $W(\phi, \xi) = V(\chi.\xi)$. It shows that the present theory actually contains, 
   in addition to $g\mn$, two dynamic degrees of freedom expressed in the scalar fields
   $\chi$ and $\xi$, or equivalently $\phi$ and $\xi$.
   
  This generalized HMPG (GHT) reduces to the initial, ``simple'' version of HMPG in the case
  $f(R, P) = R + f_1(P)$ with an arbitrary function $f_1(P)$. We then obtain $\chi =1$,
  and the scalar-tensor formulation of the theory then contains only one scalar field
  $\phi = 1 + \xi$, in agreement with \cite{har12} and all later papers on the subject.
  
  In the framework of GHT, the action \rf{S2} describes an example of a multiscalar-tensor theory 
  with two scalar fields $\phi$ and $\xi$, which, as any such theory, admits a well-known 
  transformation \cite{wag70} to the Einstein conformal frame, defined in a manifold to be denoted
  $\ME$,  in which the nonminimal coupling of the scalar field to the metric is excluded (while the 
  original formulation \rf{S2} is called the Jordan conformal frame, which is defined in the manifold $\MJ$). 
  In our case, the transformation can be written as 
\beq 			\label{J-E}
		\og\mn = \phi g\mn, \qquad    \phi = \e^{2\ophi}, \qquad   \xi = \eps \psi^2/3,
\eeq		
  and results in 
\bearr 			\label{SE}
        S_g = \int \! d^4x \sqrt{-\og}
			\bigg[\oR + 6 \og\MN \ophi_{,\mu}\ophi_{,\nu} 
\nnn \cm\			
				+ 2 \eps \e^{-2\ophi}\og\MN\psi_{,\mu}\psi_{,\nu} - 2 U(\ophi, \psi)\bigg],
\ear            
   where bars mark quantities obtained from or with the transformed metric $\og\mn$,
\beq                    \label{WU}
						U (\ophi, \psi) = \frac{W(\phi,\xi)}{\phi^2}
\eeq   
  and $\eps = \sign \xi$, so that $\chi$ (and hence $\psi$) is a canonical scalar field if 
  $\eps =+1$ and a phantom scalar field with negative kinetic energy if $ \eps =-1$. 
  Let us also note that we only consider $\phi > 0$, otherwise, as can be seen from \rf{S2},
  we would obtain a negative effective gravitational constant.
   
  Variation of the action \rf{SE} with respect to $\og\MN$, $\ophi$ and $\psi$ leads to
  the field equations
\bearr    \label{EE}
		\oG\mN = - \oT\mN \quad\ \Longleftrightarrow 
					\quad\ \oR\mN = -\oT\mN + \half \delta\mN \oT^\alpha_\alpha,
\yyy            \label{Tmn}
		\oT\mN = 6\ophi_{,\mu}\ophi^{,\nu} - 3\delta\mN (\bar\d\ophi)^2	
\nnn \quad		
		+	\big[2\psi_{,\mu}\psi^{,\nu} 
		- \delta\mN (\bar\d\psi)^2 \big] \e^{-2\ophi}	+ \delta\mN U(\ophi, \psi),	
\yyy            \label{eq-ophi}
			    6 {\bar\Box} \ophi + 2\eps \e^{-2\ophi} (\bar\d\psi)^2 + \frac{\d U}{\d\ophi} = 0,
\yyy            \label{eq-psi}
			2\eps \bar\nabla_\mu \big(\og\MN \psi_{,\nu} \e^{-2\ophi}\big) + \frac{\d U}{\d\psi} = 0.
\ear  
  
  In what follows we will try to solve the Einstein-frame field equations \rf{EE}--\rf{eq-psi} and convert 
  them to $\MJ$ with the metric $g\mn$ in which the extended HMPG theory is originally formulated.  
       
\section{Solutions for $V(\ophi, \psi) \equiv 0$} 
\subsection{Solutions in the Einstein frame}

  We begin our study with the simplest case $U(\ophi, \psi) \equiv 0$, which corresponds to 
  the original potential $V(\chi, \xi) \equiv 0$. According to \rf{VRP}, this condition means that the 
  function $f(R,P)$ in the action \rf{GS} satisfies the first-order partial differential equation
\beq         \label{V=0}
  			Rf_R + P f_P = f,
\eeq  
  whose general solution, found by standard methods, reads 
\beq           \label {f-V0}
			f(R, P) = \sqrt{RP} \Psi(R/P), 
\eeq
  where $\Psi(R/P)$ is an arbitrary function.
  It turns out that this $f(R,P)$ violates the requirement \rf{Hess}, whatever be the choice of 
  $\Psi(R/P)$, hence the scalar-tensor representation of the theory is not equivalent to the 
  original theory. However, any such function $f(R,P)$ may be interpreted ``by continuity'' as
  a member of a family of functions which, in general, satisfy \rf{Hess}, and then any solution 
  obtained with this $f(R,P)$ may be interpreted as a particular member of a family of solutions 
  corresponding to this wider choice of, in general, admissible functions $f(R,P)$. All solutions 
  discussed in this section should be understood with regard to this remark.

  Let us now find the \ssph\  solution of \eqs \rf{EE}--\rf{eq-psi} in the case $U(\ophi, \psi) \equiv 0$,
  writing the metric in the general form 
\bearr                                \label{ds_E}
	  ds^2_{\rm E} = \e^{2\gamma(u)}dt^2 - \e^{2\alpha(u)}du^2 - \e^{2\beta(u)} d\Omega^2, 
\nnn \cm
			  d\Omega^2 = d\theta^2 + \sin^2 \theta\, d \varphi^2,
\ear
  where $u$ is a radial coordinate, and assuming $\ophi = \ophi(u),\ \psi = \psi(u)$. 
  We also use the notation $r(u) \equiv \e^{\beta(u)}$ for the spherical radius.
  The stress-energy tensor (SET) $\oT\mN$ of the two scalar fields reads
\bearr                \label{SET-0}
                    \oT\mN = \e^{-2\alpha}(3 \ophi'{}^2 + \eps \psi'{}^2 \e^{-2\ophi})
\nnn \inch \times                    
                    \diag(1, -1, 1, 1),
\ear
  where the prime denotes $d/du$. 
  The algebraic structure of this SET is the same as that for a single massless minimally coupled 
  scalar field, so that the metric can be found from \eqs\rf{EE} in the same manner, as in any 
  sigma-model with zero potential \cite{chiral}. The solution is conveniently expressed in a unified
  form for both canonical and phantom behaviors of the scalars (both are possible due to $\eps = \pm 1$)
  using the harmonic radial coordinate $u$ defined by the coordinate condition \cite{kb73}
\beq         \label{harm}
		\alpha(u) = 2\beta(u) + \gamma(u),
\eeq    
  Then the combinations of \rf{EE} $\oR^0_0 =0$ and $\oG^1_1+\oG^2_2 =0$ are easily integrated
  \cite{kb73}, and the metric $ds_E^2 = \og\mn dx^\mu dx^\nu$ can be written in the form  
\beq 			\label{Fish}
            ds_E^2 = \e^{-2hu} dt^2 - \frac {\e^{2hu}}{s^2(k,u)}
            					\bigg[\frac {du^2}{s^2(k,u)} + d\Omega^2 \bigg],
\eeq      
   where 
\bearr     \label{s}
           e^{-\beta(u)-\gamma(u)} = s(k,u) := \vars {\!
                        k^{-1}\sinh ku,  \ & k > 0 \\
                                    u,  \ & k = 0 \\
                        k^{-1}\sin ku,   \ & k < 0,  }           
\nnn			
							\qquad	   h, \ k = \const.
\ear          
  Without loss of generality, the radial coordinate $u$ is defined at $u > 0$, so that $u=0$
  corresponds to spatial infinity; near it, the spherical radius $r =\sqrt{ -\og_{22}}$ behaves 
  as $r \sim 1/u$, and the Schwarzschild mass in the Einstein frame\footnote
  		{If we write the general static, spherically symmetric metric in the form \rf{ds_E} with 
  		  an arbitrary radial coordinate $u$, this metric is asymptotically flat at some 
  		  $u = u_0$ if \cite{bbook12}
 	\[
 		     \e^{\beta(u)} \equiv r(u) \toas_{u\to u_0} \infty, \quad |\gamma(u_0)| < \infty,
 		     		\quad  \e^{\beta -\alpha}|\beta'|  \toas_{u\to  u_0} 1.
	\] 		  
		Comparing \rf{ds_E} with the Schwarzschild metric, it is easy to obtain a general 
		expression for the Schwarzschild mass at $u = u_0$ :
	\[
		m = \lim\limits_{u \to u_0} \e^\beta \gamma'/\beta' = \lim\limits_{u \to u_0} r^2\gamma'/r'. 
	\]
		In particular, for the (anti-)Fisher metric \rf{Fish} we have $m = h$ at $u_0 = 0$.
		}
  is equal to $h$, while the scalar fields and 
  the two integration constants $h$ and $k$ are constrained by the relation
\beq                                                       \label{int0}
            N := k^2\sign k - h^2 = 3 \ophi'^2 + \eps \psi'^2 \e^{-2\ophi}            
\eeq
  that follows from the ${1 \choose 1}$ component of \eqs\rf{EE}.
  More detailed discussions of this metric in the context of solutions with a single 
  scalar field may be found, e.g., in \cite{h_ell73, bbook12, kb-stab11, Par18}.
  
  In the system under study, with \rf{harm}, the scalar field equations read
\bearr      \label{ophi''}
		3\ophi'' + \eps \psi'^2 \e^{-2\ophi} =0,
\yyy		\label{psi'}
		(\psi' \e^{-2\ophi})' = 0 \ \then \psi' \e^{-2\ophi} = C = \const.
\ear
  Substituting $\psi'$ from \rf{psi'} to \rf{int0}, or equivalently to \rf{ophi''} and then integrating, 
  we obtain an easily integrable equation determining $\ophi(u)$, 
\beq                   \label{int1}
           3 \ophi'^2 = N - \eps C^2 \e^{2\ophi}.
\eeq  
  The particular form of $\ophi(u)$ depends on $\eps = \pm 1$ and the values of $N$ and $C$. 
  Let us here recall that the Jordan-frame metric is $g\mn = (1/\phi) \og\mn$, where the conformal 
  factor is $1/\phi = \e^{-2\ophi}$, so its expression is of utmost importance.   
  
\subsection{Branch 1:  $\eps = +1$}

  From \rf {int0} it follows $N > 0$, in the metric \rf{s} we have $k > 0$,
  $s(k,u) = (1/k) \sinh (ku)$, and the metric \rf{Fish} corresponds to Fisher's solution \cite{fish48}
  with a canonical scalar field.
  The coordinate $u \in \R_+$, and at $u\to \infty$ we have an attracting naked central ($r\to 0$)
  singularity. The metric looks more transparent using the ``quasiglobal'' coordinate $x$ 
  defined by the condition $\alpha + \gamma =0$ in \rf{ds_E}: after the substitution
\bearr        \label{u->x}
			  \e^{-2ku} = 1 - \frac{2k}{x},\qquad \e^{-2hu} = \Big(1 - \frac{2k}{x}\Big)^a, 
\nnn \inch			  
			  a = h/k < 1,
\ear  
   the metric transforms to
\bearr                  \label{ds1}
                ds_E^2 = \Big(1 - \frac{2k}{x}\Big)^a dt^2 - \Big(1 - \frac{2k}{x}\Big)^{-a} dx^2
\nnn \inch  \ \ \              
                			  - x^2 \Big(1 - \frac{2k}{x}\Big)^{1-a} d\Omega^2 \Big],  
\ear
  However, the coordinate $u$ is still more convenient for our further consideration. 

  For the scalar fields $\ophi$, \eqn{int1} gives
\beq                   \label{u1}
		\pm du= \frac {\sqrt{3}d \ophi}{\sqrt{3 n^2 - C^2\e^{2\ophi}}}, \qquad n = \sqrt{N/3}. 
\eeq   
  Integration of \rf{u1} gives without loss of generality
\beq
		u + u_1 = \frac{1}{n} \tanh^{-1} \bigg[ \frac{N - C^2 \e^{2\ophi}}{\sqrt{N}}\bigg],
\eeq  
  with $u_1 = \const$. Solving this expression for $\e^{2\ophi}$, we find for $\e^{-2\ophi} = 1/\phi$
  (the conformal factor for a transition to Jordan's frame, $g\mn = (1/\phi) \og\mn$):
\beq               \label{phi1}
		\e^{-2\phi} = \frac {C^2}{3 n^2} \cosh^2 [n (u + u_1)].	
\eeq  
  For $\psi(u)$ we obtain from  \rf{psi'}
\beq
		\psi = \frac{3n}{C} \tanh [n(u + u_1)] + \psi_0, \quad \psi_0 = \const.		
\eeq
  
  The Jordan-frame metric has the form 
\bearr			\label{ds-J1}
            ds_J^2 =  \frac {C^2}{3 n^2} \cosh^2 [n (u + u_1)]\bigg\{	
		            		\e^{-2hu} dt^2 
\nnn \cm\ \ 		            		
		- \frac {k^2 \e^{2hu}}{\sinh^2(ku)}
            					\bigg[\frac {k^2 du^2}{\sinh^2(ku)} + d\Omega^2 \bigg]\bigg\},
\ear
  it remains \asflat\ at $u=0$ (with a changed length scale), and the \Scw\ mass is 
\beq   \label{m1}
		m_J = \frac {|C|k}{\sqrt{3}n}\big[ h \cosh (nu_1)  - n \sinh (nu_1)\big].
\eeq      		
  However, its global properties depend on the relationships between the constants since 
  at $u \to \infty$ one finds three kinds of the behavior:
\begin{description}     
	\item[1A.]
		$n < h$:\ we obtain $g_{00} \to 0$, and $r \to \infty$. It means that the radius 
		$r(u)$ has a regular minimum (a throat) at some $u>0$. beyond which, as $u\to\infty$,
		there is an attracting naked 
		singularity, a configuration sometimes called a ``space pocket'' \cite{jord}.
	\item[1B.]
		$n > h$: \ in this case, at large $u$, $g_{00} \to \infty$ and $r \to 0$, so it is a naked 
		singularity at the center, repulsive for test particles, resembling a \RN\ singularity.
	\item[1C.]
		$n =h >0$, $k = 2h$: in this case, both $g_{00}$ and $r$ have finite limits as $u = \infty$,
		so it is a regular sphere, beyond which a continuation is necessary. 
\end{description}
    Such a continuation beyond $u=\infty$ can be obtained by putting
\beq                      \label{uy} 
		    y = \coth hu, \qquad   u = \frac 1{2h} \ln \frac {y+1}{y-1}.
\eeq    
    The metric \rf{ds-J1} takes the firm
\bearr                   \label{ds-J1a}
	   ds_J^2  = \frac {C^2}{3n^2}\, \frac{(y+y_1)^2}{1-y_1^2}
	   	\bigg[\frac{dt^2}{(y+1)^2} 
\nnn \cm\qquad   	
	   			- \frac{h^2}{y^4} (y+1)^2 (dy^2 + y^2 d\Omega^2)\bigg],
\ear
  where $y_1 = \tanh (nu_1)$. The sphere $u=\infty\ \leftrightarrow\ y =1$ now becomes 
  evidently regular.  The limit $y \to \infty$ corresponds to $u\to 0$, where the metric is \asflat. 
  At the other end of the range of $y$ we have:
\begin{description}     
\item[1C(i).]
	$y_1 <  0 \ \then  y = - y_1 > 0$ is an attracting naked central singularity ($r\to 0$).
\item[1C(ii).]
	$y_1 >  0 \  \then  \ y \to 0$ is one more flat infinity, the whole space-time is thus a
	{\it traversable \wh}, \asflat\ on both ends.
\item[1C(iii).]
	$y_1 =  0  \then $ the sphere $y =0$ is a double horizon. Denoting $r = h(y+1)$, 
	we arrive at
\bearr              \label{ds-J1b}
	ds_J^2 = \frac{C^2}{3n^2} \bigg[\Big(1 - \frac hr \Big)^2 dt^2 
						- \Big(1 - \frac hr \Big)^{-2} dr^2 
\nnn \inch\ \						
						- r^2 d\Omega^2\bigg],
\ear
  which reproduces the (rescaled by $C^2/N$)  metric of a {\it \bh\ with a conformal scalar field}
  \cite {kb70, bek74}, sometimes called the BBMB \bh\ solution.	
\end{description}

  In the case $n =h >0$, the coordinate $u$ is only meaningful at $y >1$, therefore, 
  the relation \rf{phi1}, connecting the two conformal frames, also makes sense only at $y >1$.
  The range $y < 1$, emerging after conformal continuation \cite{kb01p, kb02}, corresponds to
  another similar Einstein-frame manifold, with the conformal factor $ -1/\phi$, where
\beq
               \phi = \frac{3n^2}{C^2}\,\frac{(y^2-1)(1-y_1^2)}{(y+y_1)^2}.
\eeq  
  After crossing the transition value $y=1$, the scalar field $\phi$ becomes negative, which means 
  that the region $y < 1$ is ``antigravitational,'' with a negative effective gravitational constant
  (see the action \rf{S2}), so that gravity itself actually becomes a phantom field \cite{b-star07}.  

  We conclude that in the case $V \equiv 0$, $\eps=+1$ the solutions generically possess 
  naked singularities, while \bh\ and \wh\ solutions emerge only in special cases as a result
  of conformal continuations, and, in addition, contain ``antigravitational''  regions.

\subsection{Branch 2: $\eps = -1$, $N = 3 n^2 >0$, $k >0$} 

  We have $k^2 > h^2$ due to \rf{int0}. In the Einstein frame, we again obtain 
  Fisher's metric with $s(k,u) = (1/k) \sinh (ku)$, which means that the canonical field $\ophi$ 
  stronger affects the metric than the phantom field $\psi$. As before, using \rf{u->x}, the 
  metric can be transformed to \rf{ds1} with $a < 1$, but it is more convenient to study
  the Jordan frame in terms of the coordinate $u$ (even though it is not harmonic there).
  
  Now, instead of \rf{u1}, we have 
\beq                   \label{u2}
		\pm du = \frac {\sqrt{3} d \ophi}{\sqrt{C^2 \e^{2\ophi} + 3n^2}}, 
\eeq  	
  which leads to 
\bearr             \label{phi2}
		\e^{-2\phi} = \frac {C^2}{3 n^2} \sinh^2 [n(u + u_1)], 
\nnn
		\psi = \psi_0 - \frac{3n}{C} \coth [n(u + u_1)], 	
\ear
  with $u_1, \psi_0 = \const$. The Jordan-frame metric takes the form 
\bearr 			\label{ds-J2}
            ds_J^2 =  \frac {C^2}{3 n^2} \sinh^2 [n (u + u_1)]\bigg\{\e^{-2hu} dt^2
\nnn \cm		            		
		 - \frac {k^2 \e^{2hu}}{\sinh^2(ku)}
            					\bigg[\frac {k^2 du^2}{\sinh^2(ku)} + d\Omega^2 \bigg]\bigg\},
\ear  
  and its properties depend on $u_1$. This metric is \asflat\ at $u\to 0$ only if $u_1 \ne 0$, 
  and the \Scw\ mass is
\beq   \label{m2}
		m_J = \frac {|C|k}{\sqrt{3}n}\big[ h \sinh (nu_1)  - n \cosh (nu_1)\big].
\eeq  
   We obtain the following cases:
\begin{description}     
\item[2A.]  
  If $u_1 < 0$, then $u$ ranges from 0 to $-u_1$; as before, the metric is \asflat\ at $u=0$,
  while $u = -u_1$ is a naked attracting central singularity ($g_{00}\to 0,\ r \to 0$).
\item[2B.]
  If $u_1 > 0$, $n \ne h$, then $u \in \R_+$, the metric is \asflat\ at $u=0$, while as $u\to \infty$
  the following cases are observed:
	\begin{description}  
	\item[2B(i).]  $n < h$, the large $u$ behavior is similar to item 1A.
	\item[2B(ii).]  $n > h$, the large $u$ behavior is similar to item 1B.
	\end{description}     
\item[2C.] 
    $u_1 > 0,\ n=h$: the large $u$ behavior only partly coincides with item 1C. We have again
    a continuation beyond the regular sphere $u=\infty$, implemented by the substitution \rf{uy},
    which now converts the metric to 
\bearr                    \label{ds-J2a}
	   ds_J^2  = \frac {C^2}{3n^2}\, \frac{(1+\eta y)^2}{1- \eta^2} 	\bigg[\frac{dt^2}{(y+1)^2} 
\nnn \cm   
	   			- \frac{h^2}{y^4} (y+1)^2 (dy^2 + y^2 d\Omega^2)\bigg],
\ear
    with $\eta = \tanh (n u_1)$. Since $\eta >0$, the metric describes a traversable \wh, 
    similarly to item 1C(ii), with asymptotic flatness at both $y \to \infty$ and $y \to 0$.
\item[2D.]
	 $u_1 =0$. The conformal factor $\sinh (nu) \sim u$ at small $u$ destroys there asymptotic 
	 flatness, and instead we have at $u=0$ a finite spherical radius: $r^2 = -g_{22} \to C^2/3$.
	 The metric has the asymptotic form
\beq                      \label{ds-2D}
 			ds_J^2 \approx \frac{C^2}{3} \bigg(u^2 dt^2 - \frac{du^2}{u^2} - d\Omega^2\bigg),
\eeq	  
	 clearly showing that $u=0$ is a double horizon. Beyond this horizon, at $u < 0$, we have 
	 a copy of the region $u > 0$ with the replacement $h \mapsto -h$.
	 
\qquad As $u \to \infty$, the metric behavior at $n \ne h$ is the same as in items 1A
	 (if $n < h$) and 1B (if $n > h$), that is, there are different kinds of singularities. 
\end{description}  	 

  In Branch 2D, in the case $n=h$, we have a regular sphere $u = \infty$ and a continuation 
  beyond it with the aid of \eq \rf{uy}, leading to the metric \rf{ds-J2a} with $\eta =0$. The old 
  region parametrized by $u$ maps into $y > 1$. The new region $0 < y < 1$ describes a 
  metric which is \asflat\ at $y =0$. The whole space-time in Jordan's 
  frame in this case consists of three regions: (a) the original one, with $h=n$,
  $u \in \R_+ \equiv y > 1$, (b) its continuation beyond the horizon $u=0$, that is, a copy 
  of the original region but with $h = -n$, ending with a central singularity where $g_{00} \to \infty$,
  and (c) its continuation beyond $(u = \infty) \equiv (y=1)$ ending with flat spatial infinity.
  Despite such a complex division, the global causal structure is the same as in any \asflat\
  \ssph\ space-time with two R-regions separated by a double horizon: it coincides with 
  that of the extremal \RN\ space-time.
 
\subsection{Branch 3: $\eps = -1$, $N = 0$, $k=h$} 

   It is the branch where the canonical ($\ophi$) and phantom ($\psi$) fields balance  
   each other, and the metric \rf{Fish} with $h =k$ turns into the \Scw\ metric written in terms of 
   the harmonic coordinate $u$. Its standard form is restored by putting $\e^{-2hu} = 1 - 2h/r$ and 
   a transition to the \Scw\ coordinate $r$. On the other hand, integration of \rf{u1} leads to
   the conformal factor for obtaining the Jordan-frame metric 
\bearr
		 \frac 1\phi = \e^{-2\ophi} = \frac {C^2}{3} (u +u_1)^2, \quad u_1 = \const,\quad {\rm and}
\nnn
		\psi = \frac 3{C (u + u_1)} + \psi_0, \quad \psi_0 = \const.
\ear
  The properties of $g\mn$ are easier described in terms of the $u$ coordinate:
\bearr        			\label{ds-J3}
		ds_J^2 =  \frac {C^2}{3} (u + u_1)^2 \bigg\{\e^{-2hu} dt^2 
\nnn \cm		
		- \frac {h^2}{(1-\e^{- 2hu})^2}
            					\bigg[\frac {h^2 du^2}{\sinh^2(hu)} + d\Omega^2 \bigg]\bigg\},
\ear  
   and depend on the values of $h$ and $u_1$:
\begin{description}     
\item[3A.]  
  If $u_1 < 0$, the metric behaves precisely as in the case 2A.
\item[3B.]     
  If $u_1 > 0$, the metric is \asflat\ at $u=0$ while as $u \to \infty$, we have either 
     \begin{description}     
		\item[3B(i).] $g_{00} \to 0,\ r\to \infty$ (a ``space pocket'') if $h \geq 0$, or
		\item[3B(ii).] $g_{00} \to \infty,\ r\to 0$ (a repulsive center) if $h < 0$.
\end{description}    
\item[3C.] 
   If $u_1 =0$, then near $u =0$ the metric behaves as in \rf{ds-2D} and has there a double 
   horizon, beyond which, at $u < 0$, there is a similar metric, though with $h \mapsto -h$. At 
   $u \to \infty$ the geometry is the same as in cases 2B(i) and 2B(ii) depending on the sign of $h$.
   Thus the whole space-time is a union of two conformally \Scw\ regions with different signs
   of $h$, hence with different singular asymptotics as $u \to +\infty$ and $u \to -\infty$, 
   separated by a double horizon located at $u=0$.
   If $h=0$, the geometry is symmetric with respect to the horizon $u=0$.
\end{description}    
  The \Scw\ mass for the metric \rf{ds-J3} at its flat-space asymptotic $u = 0$ is 
\beq
			m_J = \frac {|C|(1 - hu_1)}{2\sqrt 3}.
\eeq  
   
\subsection{Branch 4: $\eps = -1$, $N = - 3 n^2 < 0$, $n >0$} 

  This branch corresponds to phantom field domination, so that 
  the metric \rf{Fish} has the ``anti-Fisher'' form \cite{kb73} (first found in other coordinates in 
 \cite{b-lei57, h_ell73}) characterized either by $|h| > k \geq 0$ or by $k < 0$, see \rf{int0}. 
  Accordingly, the solution for the metric contains three families depending on the sign of $k$.
 Irrespective of the sign of $k$, \eqs \rf{int1} and \rf{psi'} lead to
\bearr                 \label{phi4}
		 \frac 1\phi = \e^{-2\ophi} = \frac {C^2}{3} \cos^2[n(u +u_1)], \quad\ u_1 = \const,
\nnn   			\label{psi4}
		\psi = \frac {3n}{C} \tan (u + u_1) + \psi_0, \qquad \psi_0 = \const.
\ear
  However, the corresponding geometries are rather diverse:

\begin{description}     
\item[4A.]  
   $k > 0$. Using again the substitution \rf{u->x}, but now with $|h|/k = a > 1$,
   we obtain the Einstein-frame metric in the same form \rf{ds1}, but its basic properties with 
   $a > 1$ are quite different from those with $a < 1$ since now $r^2_E = - \og_{22} \to \infty$
   as $x \to 2k$. However, this limit, corresponding to $u\to\infty$, is not reached in the Jordan frame
   since the conformal factor \rf{phi4} turns both $g_{00}$ and $r$ to zero at some finite $u$, 
   resulting in a central attracting singularity. The solution with
\bearr
		ds_J^2 =  \frac {C^2}{3} \cos^2[n(u + u_1)] \bigg\{\e^{-2hu} dt^2 
\nnn \quad\		
		- \frac {k^2 \e^{2hu}}{\sinh^2(ku)}
            					\bigg[\frac {k^2 du^2}{\sinh^2(ku)} + d\Omega^2 \bigg]\bigg\}
\ear  
  ranges (without loss of generality) between $u =0$ and the nearest zero of the function 
  $\cos [n(u + u_1)]$. 
\begin{description}     
		\item[4A(i).] $\cos (nu_1) \ne 0$, then the metric is \asflat\ at $u=0$, and
		the corresponding \Scw\ mass is
\beq   	\nqq		\label{mJ4A}
		m_J = \frac {|C|}{\sqrt 3} \big[ h \cos (nu_1) + n \sin (nu_1) \big].
\eeq  
		\item[4A(ii).] $\cos (nu_1) = 0$, then near $u=0$ the metric behaves as in \rf{ds-2D}, and $u=0$ is
		a double horizon. The space-time consists of two regions, each interpolating between this horizon
		at $u=0$ and a singularity at $nu = \pm \pi/2$ where $\cos (nu) =0$.
	\end{description}   
\item[4B.]   
   $k = 0$. We can now substitute $u = 1/x$ in the Einstein-frame metric, so that now $x=\infty$
    corresponds to flat spatial infinity, with the same mass \rf{mJ4A}, and $x=0$ to a singularity:
\bearr        \nq             \label{ds4b}
	  ds_E^2 =  \e^{-2h/x}dt^2 - \e^{2h/x}(dx^2 + x^2 d\Omega^2). 
\ear               
   However, the conformal factor \rf{phi4} makes the Jordan-frame metric behavior in quite a 
   similar way as that described in the cases 4A(i) and 4A(ii).
\item[4C.]   
   $k < 0$. Now the Einstein-frame metric 
\beq          \nqq                           \label{ds4c}
	ds_E^2 = \e^{-2hu}dt^2 - \frac{k^2 \e^{2hu}}{\sin^2 ku}
	                        \bigg(\frac{k^2 du^2}{\sin^2 ku} + d\Omega^2\bigg)
\eeq    
  describes a traversable \wh\ with two flat spatial infinities occurring at $u=0$ and $u = \pi/|k|$.   
  The properties of the Jordan-frame metric
\beq
			ds_J^2 =  \frac {C^2}{3} \cos^2[n(u + u_1)] ds_E^2,
\eeq  
   depend on the interplay between $\cos[n(u + u_1)]$ and $\sin(|k|u)$. 
   As before, we assume that $u \in (0, u_{\max})$, and this $u_{\max}$ depends on which of the 
   sinusoidal functions will be the first to vanish as $u$ increases. If $\cos (nu_1) \ne 0$, then
   the metric is \asflat\ at $u=0$, and the \Scw\ mass is there again determined by \eqn{mJ4A}. 
   We now have, according to \rf{int0}, $3n^2 = k^2 + h^2$. The following cases are observed:
  \begin{description}     
		\item[4C(i).] $\cos (nu_1) \ne 0$, $\cos [u_{\max} + u_1]= 0$. The geometry is similar to 
			 the cases 2A and 3A. 			
		\item[4C(ii).] $\cos (nu_1) \ne 0$, $u_{\max} = \pi/|k|$. A twice \asflat\ traversable \wh,
			with a geometry only smoothly deformed as compared to \rf{ds4c}.
			The \Scw\ mass at the other infinity, $u = u_{\max}$, is different from that at $u=0$:
	\bearr\nqq
				m_{J2} = - \frac{|C|}{\sqrt 3} \e^{hu_{\max}}
						\Big[ h \cos[n(u_1 + u_{\max})]
	\nnn \cm					
						 + n \sin[n(u_1 + u_{\max})]\Big].
	\ear		
		\item[4C(iii).] $\cos (nu_1) \ne 0$, $u_{\max} = \pi/|k|$ and simultaneously 
			$\cos [\pi/|k| + u_1]= 0$. Then $u = u_{\max}$ is a double horizon, and a 
			continuation beyond it leads to one more flat infinity at $u = 2\pi/|k|$ (since in this 
			case $n < |k|$, and the plot of $\cos[n(u + u_1)]$ is wider than that of  $\sin (ku)$).
           \item[4C(iv).] $\cos (nu_1) = 0$, $n < |k|$.   		
		 	Near $u=0$ the metric behaves as in \rf{ds-2D}, and $u=0$ is a double horizon. 
		 	The geometry is similar to the case 4C(iii), but now $u$ ranges from $-\pi/|k|$ and $\pi/|k|$,
		 	it is \asflat\ at both ends, and a horizon occurs at $u=0$.
           \item[4C(v).] $\cos (nu_1) = 0$, $n > |k|$. Again, $u=0$ is a double horizon, but now 
            		the geometry is quite similar to the case 4A(ii).  	
	      \item[4C(vi).] $\cos (nu_1) = 0$, $n = |k|$, so that $h^2 = 2n^2$, and
			$\cos [n(u+u_1)] = \sin (|k|u) = \sin (nu)$. The metric takes the form 
\bearr  \nqq                     \label{ds_inf}
		ds_J^2 = \frac{C^2}{3n^2}\sin^2 (nu) \e^{-2hu} dt^2
\nnn \quad\	
		 - n^2 \e^{2hu}	\bigg[ \frac{n^2 du^2}{\sin^2 (nu)} + d\Omega^2 \bigg].
\ear  		
    \end{description}    
 \end{description}
      
  The metric \rf{ds_inf} is defined on the whole real axis, $u \in \R$, and describes a space-time 
  that unifies an infinite number of static regions. Each of the latter corresponds to a half-wave of 
  the function $\sin (nu)$), and they are separated by double horizons occurring at each $u = \pi s/n$, 
  with any integer $s$. The Jordan-frame manifold $\MJ$ is thus constructed by conformal 
  continuations from a countable set of Einstein-frame manifolds $\ME$, each of them being a 
  \wh\ whose both infinities are mapped into horizons in $\MJ$.
 
  Concluding, we can notice that Branches 1 and 4 completely reproduce the results previously
  obtained in the framework of HMPG \cite{kb19, hyb-sph}, where they have been discussed in more 
  detail, with appropriate plots and Carter-Penrose diagrams. Branches 2 and 3 are new in 
  GHT and emerge due to the interplay of the two scalars $\ophi$ and $\psi$. Their behavior is 
  either very simple or repeats the main features of the already described solutions..  
   
 \section{Solutions for $V(\phi,\psi)\not\equiv 0$}     
\subsection{General consideration}
   In this section we are going to discuss some examples of generalized HMPG with nonzero scalar 
  field potentials. Before that, let us recall some general theorems on scalar-vacuum space-times 
  with $V \not\equiv 0$ which reveal many features of a possible behavior of the solutions 
  to the equations of motion without solving them analytically or numerically. Such theorems 
  most frequently characterize the properties of minimally coupled scalar fields, but fortunately
  concern not only systems with a single scalar field but also their multiplets or sigma models.
  Such are, for example, the no-hair theorems (for a recent review see, e.g, \cite{herd15}) 
  indicating the conditions excluding black hole horizons, and the global structure theorems 
  \cite{kb01} informing us on the existence conditions of regular solutions and on the 
  possible number of Killing horizons. Thus, it is known that with a potential $V \geq 0$, an \asflat\ 
  \bh\ with a nontrivial set of canonical (nonphantom) scalar fields is impossible \cite{ad-pier}. 
  It is also known \cite{kb01} that static, spherically symmetric space-times with any sets of minimally 
  coupled scalar fields, with any potentials, cannot contain more than two Killing horizons, and only 
  one horizon is possible in \asflat\ configurations. 

  Many of such theorems are preserved without changes for Jordan-frame space-times, conformal 
  to those with minimally coupled fields, if the conformal factor that connects them is everywhere 
  finite and regular since in this case an \asflat\ space-time remains \asflat\ at infinity, and a horizon 
  maps to a horizon, while the potential preserves its sign. The situation is different if 
  the conformal factor somewhere tends to zero or infinity. This can change the nature of 
  singularities, if any, and in some cases we meet conformal continuations. As we saw in the case 
  $V \equiv 0$, such continuations can be numerous, and as a result, the number and nature of 
  horizons in $\MJ$ may be different from that in $\ME$.
  
  It proves to be helpful to write the field equations in $\ME$ in terms of the quasiglobal radial 
  coordinate $x$, such that in the metric \rf{ds_E} $\e^{2\gamma} = \e^{-2\alpha} = A(x)$ and
  $\e^\beta = r(x)$, and we have 
\beq            \label{ds-q}
		ds_E^2 = A(x) dt^2 - \frac {dx^2}{A(x)} - r^2(x) d\Omega^2.
\eeq   
  Then, for our \ssph\ system, the components $R^0_0 = \ldots$, $G^1_1 = \ldots$ of the 
  Einstein equations \rf{EE} and their combinations ${0\choose 0}{-}{1\choose 1}$ and 
  ${0\choose 0}{-}{2\choose 2}$ can be written as
\bearr               \label{E00}
  		\frac{1}{2r^2} (A' r^2)' = - U(\ophi, \psi),
\\ \lal              \label{E11}
		\frac{1}{r^2} \big(-1 + A' rr' + Ar'{}^2\big) 
\nnn \cm		
		= A \big(\ophi'{}^2 +\eps \e^{-2\ophi} \psi'{}^2\big) - U,  	
\\ \lal              \label{E01}
		\frac{r''}{r} = - \ophi'{}^2 - \eps \e^{-2\ophi} \psi'{}^2,		
\yyy            \label{E02}
		A (r^2)'' - A'' r^2 = 2,
\ear		
  where the prime denotes $d/dx$. The scalar field equations \rf{eq-ophi} and \rf{eq-psi} read		
\bearr      	               \label{Aophi''}
	 	\frac{6}{r^2} (Ar^2 \ophi')'+ 2 \eps A \e^{-2\ophi} \psi'{}^2 - U_\ophi =0,		
\yyy	 			               \label{psi''}
	      \frac{2\eps}{r^2}(Ar^2 \e^{-2\ophi} \psi')' - U_\psi =0,	
\ear  
   where $U_\ophi = dU/d\ophi$, $U_\psi = dU/d\psi$. Among the four equations \rf{E00}--\rf{E02}
   only two are independent, in particular, the first-order equation \rf{E11} follows from \rf{E00} and 
   \rf{E02}. Also, \eqn{E02} is easily integrated giving 
\beq                       \label{I02}
			\Big(\frac A{r^2}\Big)' = \frac{2(x_0 - x)}{r^4}, \qquad   x_0 = \const. 
\eeq   
  Further, it is possible to exclude $\psi'^2$ from \eqn{Aophi''} with the aid of \rf{E01}, to obtain
\beq
		   6 (A r^2 \ophi')' - 2 A r^2 \Big(\ophi'{}^2 + \frac {r''}{r}\Big) = r^2 U_\ophi, 
\eeq  
  though, $U_\ophi$ may still contain $\psi$. 
  
  Suppose we know the metric, then from \rf{E00} we know the potential $U$ as a function of $x$,
  and we are free to suppose some kind of scalar field dependence, for example, $U = U (\ophi)$. 
  Under this assumption, \eqn{psi''} is integrated giving  
\beq                         \label{Apsi'}
                   Ar^2  \e^{-2\ophi} \psi' = C = \const;
\eeq
  substituting this into \rf{ophi''} and multiplying by $A r^4 \ophi'$, we obtain the equation
\beq
                6 A r^2 \ophi'(A r^2 \ophi')' + 2\eps C^2 \ophi' \e^{2\ophi} = Ar^4 U_\ophi \ophi'(x), 
\eeq  
  whose integration leads to a first-order equation for finding $\ophi(x)$: 
\beq                         \label{ophi'1}
   			3A^2 r^4 \ophi'{}^2 + \eps C^2\e^{2\ophi} = \int A r^4 U'(x) dx.
\eeq 
  Another, simpler first-order equation for finding $\ophi(x)$ is obtained by substituting 
  \rf{psi'} into \rf{E01}, whence
\beq                         \label{ophi'2}
   			3A^2 r^4 \ophi'{}^2 + \eps C^2 \e^{2\ophi} = -A^2 r^3 r''.
\eeq   
  For our set of equations to be consistent, \eqs \rf{ophi'1} and \rf{ophi'2} must coincide.
  This can be verified in the following way: comparing  \rf{ophi'1} and \rf{ophi'2} and 
  differentiating, we obtain 
\bearr 				\label{U'}
			U' = - \frac{(A^2 r^3 r'')'}{Ar^4} = -2A' \frac{r''}{r} 
\nnn \cm\inch		
			-3A\frac{r'r''}{r^2} - A\frac{r'''}{r}.
\ear  
  On the other hand, we have the expression for $U(x)$ given by \rf{E00}. As is directly verified,
  this expression leads to  $U'(x)$ precisely coinciding with \rf{U'}. (For this comparison, while 
  calculating $U'$, one should use \eqn{E02} to get rid of $A''$.) This proves the consistency of our
  equations.
  
  One can recall that if there is a single minimally coupled scalar field in GR, its equation follows from
  the Einstein equations and the conservation law. In our case with two scalars, we have actually
  shown that the equation for $\ophi(x)$ follows from the Einstein equations plus the equation for 
  $\psi(x)$, and this holds true under the additional assumption that the potential $U$ depends on
  $\ophi$ only.
  
\subsection{Possible behavior of the solutions}  
  
  Before considering special examples, let us outline the possible behaviors of solutions to \eqn{ophi'2}
  assuming that the metric in $\ME$ (that is, the functions $A(x)$ and $r(x)$) is known and that  
  this metric is \asflat\ as $x\to \infty$. Rewriting \eqn{ophi'2} in the form
\beq                         \label{Fi}
   			3\ophi'{}^2 = - \frac{\eps C^2 \e^{2\ophi}}{A^2 r^4 } - \frac {r''}{r},
\eeq     
  we require that the r.h.s. should be nonnegative. Taking this into account, let us try to understand, 
  what can be the left end of the range $\{x\}$ of solutions to \eqn{Fi}, if the right end is flat infinity,
  and what can be then said about the corresponding solution in Jordan's frame, $\MJ$. Recall that in  
  $\MJ$ the metric is $g\mn = (1/\phi) \og\mn$, where $\og\mn$ is given by \rf{ds-q}, and the 
  conformal factor $1/\phi = \e^{-2\ophi}$ depends on the solution of \eqn{Fi}. 
    
  As $x\to \infty$, we have $A \to 1$ and $r \approx x$, hence the whole r.h.s. of \rf{Fi}
  behaves as $x^{-4}$ if $\ophi$ has a finite limit. Such a finite limit agrees with $\ophi' \sim x^{-2}$,
  so this is a generic behavior. Then the conformal factor $\e^{-2\ophi}$ is also finite at large $x$,
  and the \asflat\ metric in $\ME$ maps to an \asflat\ metric in $\MJ$.
  
  There is one more special opportunity in the case $\eps = -1$, that $\ophi \approx \log x$ as 
  $x\to \infty$, in which case $\ophi'{}^2 \sim \e^{2\ophi}/x^4 \sim x^{-2}$, and there is no 
  asymptotic flatness in $\MJ$. This option will not be considered. 
  
  The other end of the $x$ range in $\ME$ can represent:
\begin{description}  
\item[(i)]
	  A space-time singularity in $\ME$ (most probably but not necessarily connected with $r=0$) 
	  due to the properties of the potential $U(\ophi)$. Such singularities can be quite diverse 
	  in nature, and it does not seem possible to describe the behavior of $\ophi$ unambiguously.	   
\item[(ii)]
	  A singularity of the form $\ophi \to \infty$ at some regular point of the metric, $x=x_s$, in $\ME$,
	  where both $A$ and $r$ are finite. An analysis of \eqn{Fi} shows that in this case
	  a solution for $\e^{-\ophi}$ is a linear function of $x$, and
   \beq 				\label{sing}  
   			\e^{-2\ophi}  \sim |x-x_s|^2  \ \ \ {\rm near} \ \ \ x=x_s. 
   \eeq	  	  
       Thus a regular sphere $x=x_s$ in $\ME$ maps to a central (such that the Jordan-frame 
       spherical radius $r_J \to 0$), attracting ($g_{tt} \to 0$) singularity in $\MJ$. 
\item[(iii)]
        A regular center $x =x_c$ in $\ME$, where $ A \to 1$ and $r(x)\approx  x-x_c$. A possible
        solution for $\ophi$ behaves there as $\const/(x-x_c)$, and for our conformal factor we have
   \beq 				\label{cen}  
			\e^{-2\ophi}  \sim \e^{-k^2/(x-x_c)} \to 0  \ \ \ {\rm as} \ \ \ x \to x_c, 
   \eeq        
       with $k= \const$. 
       We conclude that a regular center in $\ME$ maps to a central attracting singularity in $\MJ$.  
\item[(iv)]
       A horizon at some $x=x_h$ in $\ME$, such that $r > 0$ and $A \sim x\!-\!x_h$. Then \eqn{Fi}
       leads to 
\bearr  				\label{hor}  
 		\ophi'{}^2 \sim \frac{\e^{2\ophi}}{(x-x_h)^2} \ \ 
\nnn 		
 		\then \ \   \e^{-2\ophi} \sim [\ln (x-x_h)]^2 \to \infty.
\ear      
      Thus a horizon in $\ME$ maps to a singularity in $\MJ$, at which the spherical radius 
      $r_J \sim [\ln (x-x_h)]^2 \to \infty$ while $g_{tt} \sim (x-x_h) [\ln (x-x_h)]^2 \to 0$, so
      the singularity is attracting.
\item[(v)]
      If $\eps = -1$, there can be no center in $\ME$ and $x \in \R$. In particular, we can have
      a twice \asflat\ \wh, with a second flat infinity at $x = -\infty$. Quite similarly to $x \to \infty$,
      we must have there generically $\ophi \to \const$, hence there will be also flat infinity in $\MJ$,
      and a twice \asflat\ \wh\ there as well.      
\item[(vi)]
       Another opportunity of interest at $\eps =-1$ is an AdS asymptotic behavior in $\ME$ 
       as $x \to -\infty$, where $r(x) \sim x$ and $A(x) \sim x^2$. 
       There can be two generic cases in \eqn{Fi}: 
     \begin{description}  
	\item[(vi-a)]
		The second term $\sim x^{-4}$ in the r.h.s. is dominant, then \rf{Fi} asymptotically gives 
	\bearr      \nqq         \label{vi-a}
		  \ophi'{}^2 \sim x^{-4} \  \then \ \ophi \sim 1/x + \const \to \const  
	\nnn \cm	  
		  {\rm as} \ \ x \to -\infty,
	\ear
          and $\MJ$ has the same asymptotic behavior as $\ME$.
      \item[(vi-b)] 
          If $\e^{2\ophi}$ grows more rapidly than $x^4$, then the first term in the r.h.s. of \rf{Fi}
          is dominant, and we have
      \beq       \nqq             \label{vi-b}
    		\ophi'{}^2 \sim \frac{\e^{2\ophi}}{x^8} \ \ \then\ \ \e^{-\ophi} \sim |x|^{-3} + C_1,
     \eeq  
         with $C_1 = \const$. Since $\e^{2\ophi}$ must grow as $x \to-\infty$, we have to put
         $C_1=0$. We then obtain $\e^{2\ophi} \sim x^6$, as required, while our conformal 
         factor $\e^{-2\ophi}$ decays as $x^{-6}$. It means that the AdS infinity in $\ME$ maps 
         to a central attracting singularity in $\MJ$.       
     \end{description}
\item[(vii)]     
    The solutions to \eqn{Fi} are not necessary monotonic. At a possible regular extremum $x=x_0$ 
    of the function $\ophi(x)$, the r.h.s. of \rf{Fi} should vanish together with its derivative in $x$, 
    since the l.h.s. is generically $\sim (x-x_0)^2$. However, such solutions to \rf{Fi} do not lead to
    valid solutions of GHT because in such cases the potential $U(x)$ known from \eqn{E00} 
    cannot be transformed to a function of $\ophi$: indeed, the same value of $\ophi$ then 
    corresponds to at least two values of $x$.    
\end{description}
    We will see how these opportunities are implemented in some particular examples.    

\subsection{Examples} 
 
 \paragraph{Example 1: A solution with canonical behavior.}    
\begin{figure*}    
\centering
\includegraphics[width=6cm]{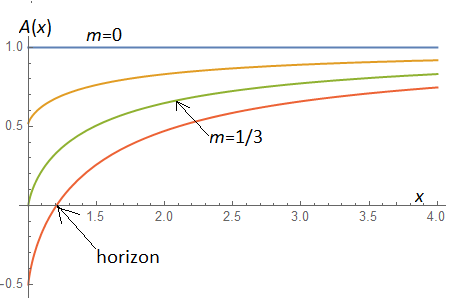}          
\qquad
\includegraphics[width=6cm]{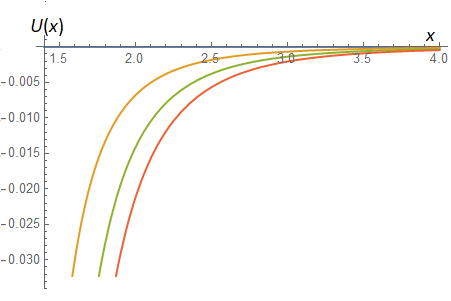}          
\caption{\small
         Example 1, the function $A(x)$ (left panel) and the potential $U(\phi(x))$ (right panel)
         for the Einstein-frame solution \rf{ds-q}, \rf{rE1}--\rf{UE1}, $m = 1/6, 1/3, 1/2$
         (upside down; at $m=0$ we have $A\equiv 1$ and $U \equiv 0$). 
         At $x = 1$, there is a naked singularity for $m \leq 1/3$ and a
         singularity beyond a horizon (at which $A(x)=0$) for $m > 1/3$.
         }
\end{figure*}    
  
  Let us try to use a known analytic scalar-vacuum solution of GR with a minimally coupled
  scalar field \cite{vac2} and the scheme outlined above to obtain a solution of GHT. 
  
  If we specify the function $r(x)$, we directly find $A(x)$ from \rf{I02}, so that the metric 
  is known completely, then \rf{E00} yields $U(x)$, and it remains to find the scalar fields
  using \rf{ophi''} and \rf{psi''}. If, as before, we suppose $U = U(\ophi)$, the field $\ophi$ 
  can be found from the first-order equation \rf{Fi}.   
 
  Following \cite{vac2}, let us choose a particular dependence $r(x)$, 
\beq                           \label{rE1}
	r(x) = \sqrt{x^2 - a^2}, 
\eeq
  and put $a=1$ as an arbitrary length scale. The inequality $r''/r = - (x^2 - 1)^{-2} < 0$ 
  confirms the canonical nature of our set of scalar fields (see \eqn{E01}). Then, from 
  \rf{I02} and \rf{E00} we find the metric function $A(x)$ and the potential \cite{vac2}
  under the assumption that the metric is \asflat, whence $A\to 1$, as $x \to \infty$:
\bearr    	                     \label{AE1}
	  A(x) = 1 - 3mx + \frac 32 m (x^2 - 1) \log \frac{x+1}{x-1},
\yyy				        \label{UE1}
	  U(x) = \frac{3m}{2(x^2\! - \!1)} \bigg[ 6x - (3x^2\! - \!1) \log \frac{x+1}{x-1}\bigg],
\ear         
  where $m= \const$ has the meaning of the Schwarzschild mass in $\ME$.
    
  The value $x = 1$ corresponds to $r=0$, which is a central naked singularity for $m \leq 1/3$ and 
  a singularity beyond an event horizon for $m > 1/3$, see Fig.\,1. The potential $U$ rapidly
  ($U \sim x^{-5}$) decays at large $x$, and $U \to -\infty$ as $x \to 1$. It is everywhere negative,
  so the existence of a horizon does not contradict the known no-hair theorems \cite{ad-pier}. 
  
  Assuming, as suggested above, $U= U(\ophi)$, we obtain \eqn{Fi} in the form
\bearr                    \label{Fi1}
              3 \ophi'{}^2 = \frac{1}{(x^2 - 1)^2} 
              	\bigg[1 - \frac{\eps C^2 \e^{2\ophi}}{A^2 (x) }\bigg],      	
\ear  
  with $A(x)$ given in \rf{AE1}. This equation can probably be solved only numerically. 
  An exception is the case $C=0$ corresponding to $\psi = \const$ which is of no interest for
  the theory under study since it leads to $\chi = \d f/\d P =0$ in its original formulation \rf{GS}
  (see \rf{scalars} and \rf{J-E}). Assuming $C \ne 0$, we notice that a change of its value is 
  equivalent to adding a constant to $\ophi(x)$, which does not affect the qualitative behavior 
  of $\e^\ophi$, and we can safely put $C=1$.  

  Next, \eqn{Fi1} may be considered with both $\eps =1$ (then both scalars $\ophi$
  and $\psi$ are canonical) and $\eps = -1$ (then $\psi$ is phantom but $\ophi$ is dominant
  at all $x$). Also, different solutions to \eqn{Fi1} correspond to different signs of $\ophi'$,
  hence there can be as many as four different solutions with the same boundary 
  value of $\ophi$ specified at some fixed value of $x$.
  
  Which kind of solutions for $\ophi(x)$ can be expected?
    
   As $x\to \infty$, \eqn{Fi1} takes the approximate form $3 x^4\ophi'{}^2 = 1 - \eps \e^{2\ophi}$. 
  This equation is easily solved by substituting $x = 1/\xi$, and the solution can be written as
\bearr           		\label{f+inf}
		\e^{-2\ophi}\Big|_{\eps=1} = \cosh^2 \bigg(C_1 \pm \frac 1 {\sqrt 3 x}\bigg),
\nnn
		\e^{-2\ophi}\Big|_{\eps=-1} = \sinh^2 \bigg(C_1 \pm \frac 1 {\sqrt 3 x}\bigg),		
\ear
  with $C_1 = \const$, so that $\e^{-2\ophi}$ has a finite limit at large $x$ with any $C_1$ if 
  $\eps =1$ and $C_1\ne 0$ if $\eps = -1$. It confirms the general observation of Sec.\,4.2.
  
  For $m \leq 1/3$, an analysis of \eqn{Fi1} shows that its solutions near the singularity $x=1$
  either tend to a finite value or (if $m=1/3$) lead to  $\e^{-2\ophi} \to \infty$, so that the 
  singularity in $\ME$ maps according to \rf{J-E} to a singularity in $\MJ$. 
  
  For $m >1/3$ we arrive at a horizon with mode (iv) of the solution behavior.
  
  In addition, for some solutions of \rf{Fi1}, as follows from our general consideration, we can expect 
  modes (ii) and (vii).
  
  These inferences are confirmed by solving \eqn{Fi} numerically, some special solutions 
  are plotted in Figs.\,1--3, where particular asymptotic values of $\ophi$ as $x\to \infty$ are 
  chosen for convenience, with no effect on the qualitative behavior of the solution.
  
  From Fig.\,2 it follows that with $\eps=+1$ at $m < 1/3$ we have either $\e^{-2\ophi} \to \infty$, as
  predicted, or we come across a monotonicity loss (mode (vii)). At $m \geq 1/3$ only mode (vii) is
  present.
  
  If $\eps = -1$ (Fig.\,3), in the case of growing $\ophi$ (left panel), we obtain a finite limit of $\ophi$ 
  at $m \leq 1/3$ (so that the singularity in $\ME$ is preserved in $\MJ$) and mode (iv), corresponding
  to a horizon, at $m > 1/3$. In the case of falling $\ophi$, hence growing $\e^{-2\ophi}$, the left ends
  of all curves correspond to mode (ii).
   
\begin{figure*}    
\centering
\includegraphics[width=6.5cm]{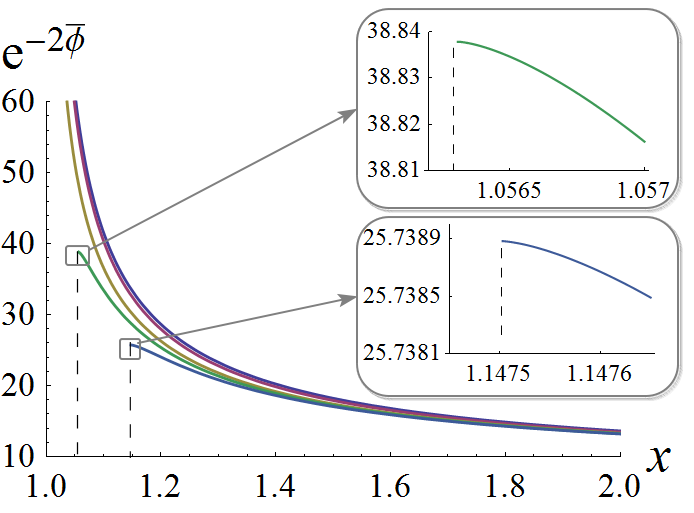}          
   \qquad
\includegraphics[width=6.5cm]{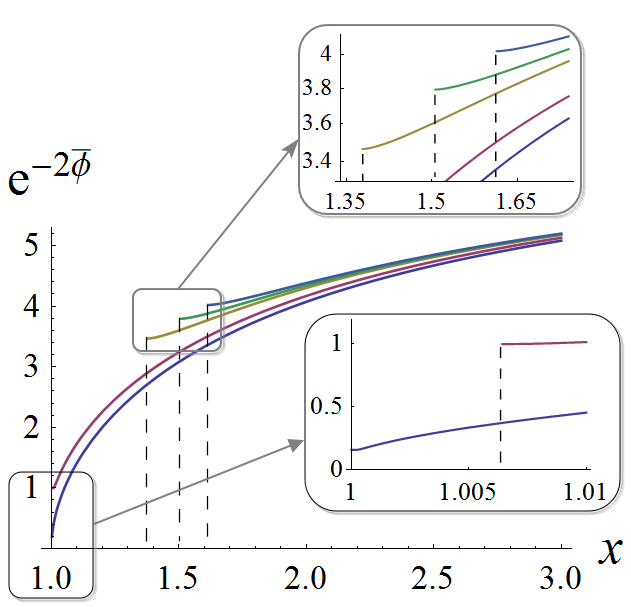}   
\caption{\small
            Example 1, the function $\e^{-2\bar\phi}$ (the conformal factor at a transition to the
            Jordan frame) for two branches of the solution to \eqn{Fi} 
            with $\eps=+1$ corresponding to $\ophi' >0$ (left panel, upside-down:
            $m = -0.5, 0, 0.28, 1/3, 0.37$), and $\ophi'<0$ (right panel, bottom-up: 
            $m = -0.5, 0, 0.28, 1/3, 0.37$). The asymptotic value is $\ophi(\infty)=-1$.
            The insets show more clearly the behavior of some curves near their termination.}
   \end{figure*}    
\begin{figure*}    
\centering
\includegraphics[width=6.5cm]{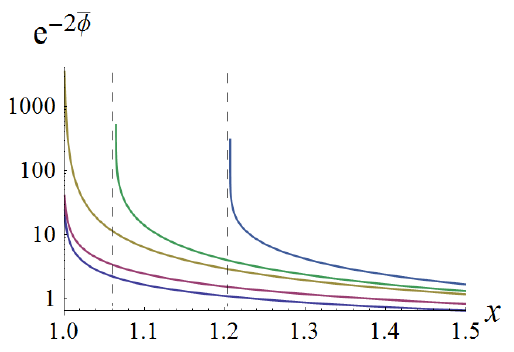}
\qquad
\includegraphics[width=6.5cm]{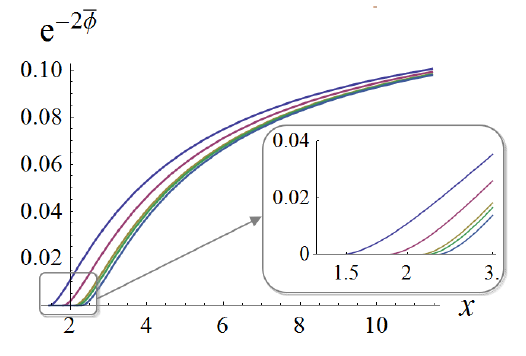}
\caption{\small
             Example 1, the function $\e^{-2\bar\phi}$ for two branches of solution to \eqn{Fi} with 
             $\eps=-1$ corresponding to $\ophi' > 0$ (left panel, bottom-up: $m = -0.5, 0, 1/3, 0.4, 0.5$),
             and $\ophi'<0$ (right panel, upside-down: $m = -0.5, 0, 1/3, 0.4, 0.5$). The asymptotic
             value is $\ophi(\infty)=1$.}
     \end{figure*} 
 
\paragraph{Example 2: A solution with phantom behavior.}    
\begin{figure*}
\centering
\includegraphics[width=6cm]{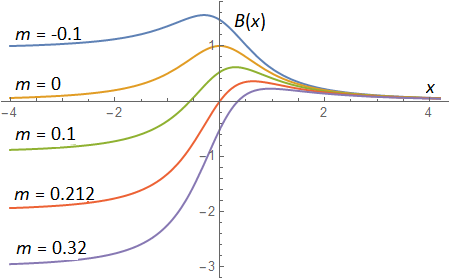}          
\qquad
\includegraphics[width=6cm]{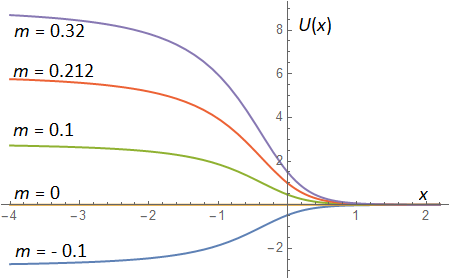}          
\caption{\small
         Example 2, the function $B(x) = A(x)/r^2(x)$ (left panel) and the potential $U(\phi(x))$ (right panel)
         for the Einstein-frame solution \rf{ds-q}, \rf{rE2}--\rf{UE2}, $m = -0.1, 0, 0.1, 0.212, 0.32$
         (bottom-up for $B(x)$, upside down for $U(x)$. As $x \to -\infty$, the metric is 
         asymptotically de Sitter if $m >0$ and AdS if $m < 0$. At $m \approx 0.212$, the horizon 
         occurs at $x=0$ and coincides with the minimum of $r(x)$.}
\end{figure*}    
\begin{figure*}    
\centering
\includegraphics[width=6.5cm]{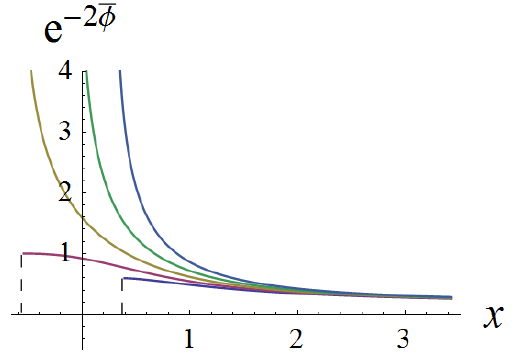}
\qquad
\includegraphics[width=6.5cm]{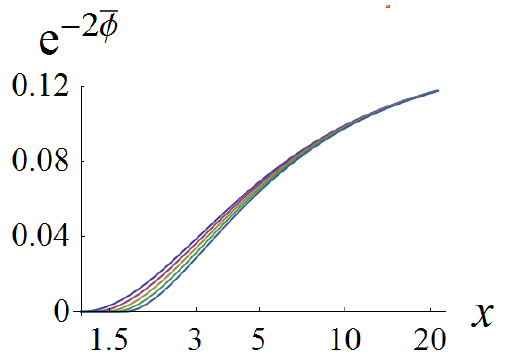}
\caption{\small
            Example 2, the function $\e^{-2\bar\phi}$ for two branches of the solution to \eqn{Fi} 		
            corresponding to $\ophi'>0$ (left panel, bottom-up: $m = -0.1, 0, 0.1, 0.2, 0.3$) 
            and $\ophi'<0$ (right panel, upside-down: $m = -0.1, 0, 0.1, 0.2, 0.3$). The asymptotic 
            value is $\ophi(\infty)=1$.}
\end{figure*}    
  In our second example let us assume, following \cite{pha1}, 
\beq                           \label{rE2}
	r(x) = \sqrt{x^2 +a^2}, 
\eeq
  and again put $a=1$ as an arbitrary length scale. The inequality $r''/r = (x^2 + 1)^{-2} > 0$ 
  confirms the phantom nature of our set of scalar fields, so that $\eps = -1$. Then, 
  assuming that the metric is \asflat\ as $x \to \infty$, we find from \rf{I02} and \rf{E00} 
  the metric function $A(x)$ and the potential $U(x)$ as \cite{pha1} as
\bearr    	                     \label{AE2}
	  A(x) = 1 + 3mx - 3m (x^2+ 1) \arccot x ,	
\yyy				        \label{UE2}
	  U(x) = \frac{3m}{(x^2+1)} \big[ -3x + (3x^2+1) \arccot x \big]. 
\ear         
  In this solution, $x \in \R$, and $m= \const$ has the meaning of the Schwarzschild mass
   in $\ME$. The behavior of the solution as $x \to -\infty$ depends on the sign of $m$:

\begin{itemize}
     \item
     		$m < 0:$\  $A \sim x^2,\ U \to \const < 0$ --- the solution describes a wormhole with an 			
     		AdS limit at the ``far end''.
     \item
     		$m=0:$  \  $A \to 1,\ U \equiv 0$ --- it is the simplest (Ellis) twice \asflat\ \wh\ with 
     		zero mass \cite{kb73, h_ell73} ($A \equiv 1$, $U \equiv 0$).	  
     \item
     		$m > 0:$ \ $A \sim -x^2, \ U \to \const >0$ --- we obtain a regular \bh\ with a de Sitter 
     		expansion far beyond the horizon instead of a singularity	(a ``{\it black universe}'' 
     		\cite{pha1, pha2}). 
\end{itemize} 
   The behavior of $B(x) \equiv r^2(x)A(x)$ and $U(x)$ is shown in Fig.\,5.

   Assuming, as before, $U= U(\ophi)$, we obtain \eqn{Fi} in the form
\bearr                    \label{Fi2}
              3 \ophi'{}^2 = \frac{1}{(x^2+1)^2} 
              	\bigg[\frac{C^2 \e^{2\ophi}}{A^2 (x) } - 1\bigg],      	
\ear  
  with $A(x)$ given in \rf{AE2}. We notice that it is necessary to put $C\ne 0$, since at 
  $C=0$ \eqn{Fi2} has no solution. With $C \ne 0$, as in Example 1, we put $C=1$ without
  loss of generality.

\medskip  
   As $x\to \infty$, \eqn{Fi2} takes the approximate form $3x^4 \ophi'{}^2 = \e^{2\ophi}-1$. 
  This equation is easily solved by substituting $x = 1/\xi$, and the solution can be written as
\beq            		\label{finf}
		\e^{-2\ophi} = \cos^2 \Big(C_2 \pm \frac 1 {\sqrt 3 x}\Big),\qquad C_2 = \const,
\eeq  
  so that $\e^{-2\ophi}$ has a finite limit at large $x$, unless $C_2 = (n+\half)\pi$ with integer $n$.
  It confirms the general observation of Sec.\,4.2.

  The other end of the $x$ range depends on the sign of $m$. The case $m=0$ returns 
  us to the already considered systems with zero potential, so let us suppose $m \ne 0$.   
  
  If $m > 0$, there is a horizon in $\ME$ at some finite $x =x_h$, where we find the behavior 
  denoted as (iv).
  
  If $m < 0$, the other end is at $x \to -\infty$, where $A(x) \sim x^2$, and we can expect 
  modes (vi-a) and (vi-b).
  
  In addition, with any $m$, we cannot exclude the emergence of modes (ii) and (vii).  
  The particular kind of solution should depend on the initial condition for $\ophi$,  
  specified at some value of $x$.
  
   Solving \eqn{Fi2} numerically, we find for $\ophi' > 0$ (Fig.\,5, left) two modes, (vii) if 
   $m \leq 0$ and (iv) if $m >0$, the latter corresponding to a horizon. For $\ophi' < 0$ 
   (Fig.\,5, right) we find mode (ii) at all values of $m$.   

\paragraph{Example 3: A trapped-ghost solution.}    
\begin{figure*}
\centering
\includegraphics[width=6cm]{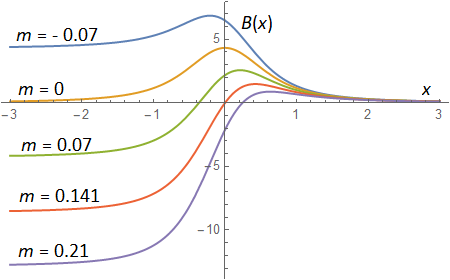}
\includegraphics[width=8cm]{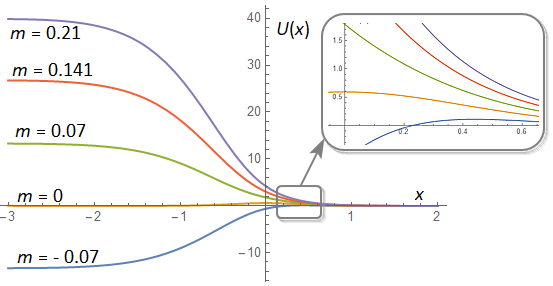}  
\caption{\small
	    Example 3, the function $B(x) = A(x)/r^2(x)$ (left panel) and the potential $U(\phi(x))$ (right panel)
         for the Einstein-frame solution \rf{ds-q}, \rf{rE3}--\rf{UE3}, $m = -0.07, 0, 0.07, 0.141, 0.21$
         (bottom-up for $B(x)$, upside down for $U(x)$). The right panel shows a detailed behavior of
         $U$ at small $x$. As $x \to -\infty$, the metric is asymptotically de Sitter if $m >0$ and 
         AdS if $m < 0$. At $m \approx 0.141$, the horizon occurs at $x=0$ and coincides with  
         the minimum of $r(x)$.}
\end{figure*}
\begin{figure*}    
\centering
\includegraphics[width=6.5cm]{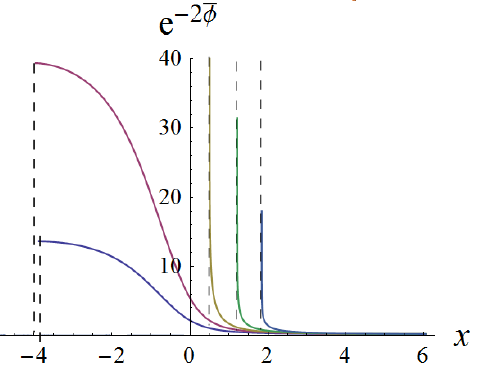}
\qquad
\includegraphics[width=6.5cm]{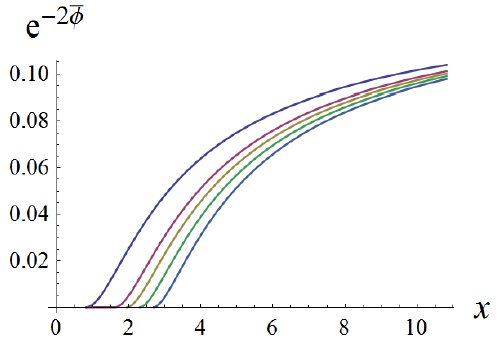}
\caption{\small
            Example 3, the function $\e^{-2\ophi}$ for two branches of solution to \eqn{Fi} corresponding 
            to $\ophi'>0$ (left panel, from left to right: $m = -0.9, 0, 0.3, 0.6, 0.9$) and $\ophi'<0$ 
            (right panel, from top to bottom: $m = -0.9, 0, 0.3, 0.6, 0.9$). 
            The asymptotic value $\ophi(\infty)=1$.}
\end{figure*}
  The following choice of $r(x)$ \cite{Par18} instead of \rf{rE1} or \rf{rE2},
\beq                           \label{rE3}
	r(x) = a\frac{x^2 + 1}{\sqrt{x^2 +3}}, \qquad a = \const,
\eeq  
  presents an example of a so-called trapped-ghost behavior \cite{trap1} of the scalar fields
  since the quantity 
\beq                            \label{r''E3}
  			\frac{r''}{r} = \frac 1{a^2} \frac{15-x^2}{(x^2 +3)^2 (x^2+1)}
\eeq  
  is positive at small $x$ (in the ``strong field region'') and negative at large $x$. Meanwhile
  (see \eqn{E01}), $r''/r > 0$ is an indicator of a phantom behavior of matter in GR because
  $r''/r \sim - T^t_t + T^x_x = - (\rho + p_r) >0$ in standard notations, which manifests NEC 
  violation. With \rf{r''E3}, we have a phantom behavior only at $x^2 < 15$.
  
  In our case with two scalars $\ophi$ and $\psi$, the first one is always canonical, and 
  the ansatz \rf{rE2} means that the second, phantom field $\psi$ is dominant at small $x$ 
  and subdominant at large $x$. 
  
  Let us put, as before, $a=1$ and require asymptotic flatness as $x \to \infty$. Then, with  
  the ansatz \rf{rE3}, the metric function $A(x)$ and the potential $U(x)$ have the form
  \cite{Par18}
\bearr         \nhq                      \label{AE3}
		A(x) = \frac{(x^2+1)^2}{x^2+3} \biggl[ \frac{39 m}{2}\Big(\arctan x - \frac \pi 2\Big)
\nnn 
		 + \frac{26 + \!24x^2\! +\! 6x^4 + 3mx(69 + 100 x^2 + 39 x^4)}{6(1 + x^2)^3}\bigg],
\nnn		 
\yyy       \nhq                           \label{UE3}
	     U(x) =	\frac 1 {12(1 +\! x^2)^2 (3 +\! x^2)^3} \Big[32(- 6 + x^2 + 3x^4)
\nnn      \nhq
		+ 6mx(2655 +\! 6930 x^2\! +\! 5420 x^4\! +\! 1326 x^6\! + \!117 x^8)
\nnn 		
		+ 117m (1 + x^2)^2 (15 + 89 x^2 + 29 x^4 + 3 x^6)
\nnn \inch \times		
		(-\pi + 2 \arctan x)\Big].     
\ear

   Despite different analytical expressions, the qualitative features of $A(x)$ and $U(x)$ are here 
   almost the same as in Example 2, see Fig.\,7. The main difference is that at $m=0$ 
   we now have $A(x) \ne \const$ and $U\not\equiv 0$, so this case (a twice \asflat\ \wh)
   is not covered by the description in Section 3.
   
   Accordingly, the behavior of the corresponding solutions to \eqn{Fi} in this case should 
   contain the same modes as in Example 2, and, in addition, mode (v) for $m=0$.   
   
   These inferences are confirmed by solving \eqn{Fi} numerically, as demonstrated by Fig.\,7.
 
 \section{Concluding remarks} 
      
  We have considered exact analytical \ssph\ solutions of GHT (generalized HMPG) 
  without matter, making use its scalar-tensor representation with two effective scalar fields
  \cite{boh13}, with zero potential $V (\chi, \xi) = W(\phi, \xi)$ and some special cases of
  nonzero potentials. The results are compared with their counterparts in the simpler version 
  of the theory, ``genuine'' HMPG \cite{har12}, obtained previously \cite{kb19,hyb-sph}.
  As before, we use the Einstein conformal frame $\ME$ to solve the equations and then
  a transition back to the original Jordan frame $\MJ$ in which the theory is formulated
  and interpreted.
  
  In the case of zero potential, $V \equiv 0$, the set of resulting space-time metrics includes
  all metrics discussed in \cite{kb19,hyb-sph} for $V \equiv 0$, plus two new families 
  (Branches 2 and 3) whose emergence is directly related with a more complex scalar 
  field configuration. While in HMPG the whole set of solutions splits into two sectors, the
  canonical and phantom ones according to the nature of the single scalar field, in the 
  present case emerges a more intricate interplay between two scalars when one of them
  is canonical and the other phantom. One of the new families of solutions corresponds 
  to their equilibrium, with the \Scw\ metric in $\ME$ and its certain deformation in $\MJ$.
  
  As in HMPG, generic solutions either contain naked singularities or, in the case of a phantom
  behavior of the scalar field set, describe traversable wormholes. Black-hole solutions 
  in $\MJ$ emerge only due to conformal continuations \cite{kb02} and form two special 
  families with double (extreme) horizons. This feature of scalar-tensor solutions with $V=0$
  could be expected due to the well-known no-hair theorems.
  
  Some exact solutions with $V \not\equiv 0$ have been obtained under assumptions 
  quite similar to those used in \cite{kb19,hyb-sph}, and the Einstein-frame metrics are the 
  same as were known previously in \cite{vac2, pha1, pha2}. However, the conformal factor 
  needed to transfer the metric to $\MJ$ is found only numerically, and it has turned out that
  even \bh\ solutions existing in $\ME$ become those with naked singularities in $\MJ$. 
  A new feature of GHT as compared to GMPG is the existence of solutions with 
  trapped-ghost properties \cite{trap1}, but in the example considered here such a solution 
  behaves similarly to the one with a phantom-like scalar set.
  
  From our Examples 1--3 one might conclude that all GHT solutions with a nonzero potential 
  contain naked singularities, since even a horizon in $\ME$ maps to a singularity in $\MJ$.
  These are, however, only special cases under the special assumption $U=U(\ophi)$, and 
  more general solutions are yet to be found.

  In \cite{hyb-sph} some results were reported on the stability properties of HMPG solutions,
  but they cannot be extended to GHT even in cases where the metrics are the same, 
  due to a more complex nature of the scalar field set. Such a stability study is a task 
  for the near future and can be probably performed by analogy with 
  \cite{b-kh79, ggs1, kb-stab11, kb-stab12, Par18} etc. Another possible continuation
  of the present study is its extension including electromagnetic fields, which can be 
  done rather easily for the case $V=0$ by analogy with \cite{kb73, CBH3} but 
  only with the aid of numerical methods for $V \ne 0$ even if we use in $\ME$ siutable 
  solutions known in GR \cite{we12,kor15} as a basis.
    
\subsection*{Funding}

 	This publication was supported by the RUDN University Strategic Academic 
	Leadership Program and by RFBR Project 19-02-00346. 
	K.B. was also funded by the Ministry of Science and Higher Education 
	of the Russian Federation, Project ``Fundamental properties of elementary 
	particles and cosmology'' N 0723-2020-0041, 
	

\end{document}